\documentclass[prd,twocolumn,
showpacs,preprintnumbers,nofootinbib]{revtex4}
\usepackage[dvips]{graphicx}
\usepackage{enumerate}
\usepackage{amsmath,amssymb}
\usepackage{mathrsfs}
\usepackage[dvips]{graphicx}
\usepackage{bm}

\begin{document}
\preprint{CHIBA-EP-188, 2011}

\title{A  low-energy effective Yang-Mills theory for quark and gluon confinement}


\author{Kei-Ichi Kondo$^{1}$}

\affiliation{$^1$Department of Physics,  
Graduate School of Science, 
Chiba University, Chiba 263-8522, Japan
}
\begin{abstract}

We derive a gauge-invariant low-energy effective model of the  Yang-Mills theory.  We find that the effective gluon propagator belongs to the Gribov-Stingl type and agrees with it when  a mass term which breaks nilpotency of the BRST symmetry is included. 
We show that the effective model with gluon propagator of the Gribov-Stingl type exhibits both quark and gluon confinement: the Wilson loop average has the area law and the Schwinger function violates  reflection positivity.
However, we argue that both quark and gluon confinement can be obtained even in the absence of such a mass term.

\end{abstract}

\pacs{12.38.Aw, 21.65.Qr}

\maketitle

It is well-known that the area law of the Wilson loop average is a gauge-invariant criterion for quark confinement.
However, a gauge-invariant criterion for gluon confinement and color confinement is not yet achieved.
In recent several years, nevertheless, great endeavors have been made to clarify the deep infrared behavior of propagators for  gluon and the Faddeev-Popov (FP) ghost in specific gauges, e.g., Landau, Coulomb and Maximally Abelian (MA) gauges. 
This research is motivated from a hope that  color confinement might be attributed to the deep infrared behavior of the gluon and ghost propagators \cite{KO79,GZ78}.

In the most common Landau gauge, especially,  it is still under debate to discriminate two different types of propagators, i.e., 
scaling \cite{AvS01} 
(an infrared suppressed gluon propagator with a finite or even vanishing dressing function at zero momentum, and a  ghost propagator more divergent in the infrared than its tree-level counterpart)
and decoupling \cite{decoupling}
(an infrared finite gluon propagator and an ghost propagator  with a finite dressing function at zero momentum) 
The scaling solution is in accordance with the Kugo-Ojima (KO) color confinement criterion \cite{KO79} and the Gribov-Zwanziger (GZ) confinement scenario \cite{GZ78}. 
On the other hand, the decoupling solution is supported by recent results of numerical simulations on the lattice with very large volumes \cite{decoupling-lattice}.  
See e.g. \cite{QCD-TNT09,Ghent-QCD} on the present status of development.

It is demonstrated \cite{FMP09}  that there is a one-parameter family of solutions for the ghost and gluon propagators of Landau gauge Yang-Mills theory and that it is only a matter of infrared boundary conditions whether infrared scaling or decoupling occurs.
Here the scaling solution is the only one member of this family that satisfies the KO/GZ property with a globally well-defined BRST charge.
The remaining solutions are of a decoupling type and cannot maintain global color symmetry and Becchi-Rouet-Stora-Tyutin (BRST) symmetry simultaneously.
Moreover, both type of solutions   violates  the reflection positivity  which is a necessary condition for gluon confinement, suggesting that neither type of solution can be associated with a massive gluon characterized by a gauge-independent pole mass. 

On the other hand,  both type of solutions in the Landau gauge has been shown to satisfy quark confinement criterion \cite{BGP10}. 
However, this result is valid for \textit{non-zero} temperature $T$ below the deconfinement temperature $T_c$ ($0<T<T_c$), 
since  vanishing Polyakov loop average at finite temperature was used as a gauge-invariant criterion for quark confinement \cite{MP08,Kondo10}. 

In this paper we examine whether there is a specific infrared behavior of the gluon propagator which is compatible with both quark confinement and gluon confinement at \textit{zero temperature}, irrespective of the gauge choice. 
For this purpose, we derive a gauge-invariant low-energy effective model of the Yang-Mills theory at zero temperature in the gauge-independent manner.
We show that the resulting low-energy effective gluon propagator belongs to the  Gribov-Stingl type \cite{Stingl86} in the low-energy region. 
In the MA gauge, especially,  the effective model is confining in the sense that the Wilson loop average has the area law and that the gluon Schwinger function violates the reflection positivity. 
In our model, an effective gluon propagator agrees with the Gribov-Stingl form  only when one includes a certain mass term violating the nilpotent BRST symmetry. 
However, we argue that such a mass term is not necessarily  indispensable to obtain quark and gluon confinement simultaneously, since both the area law and positivity violation can be obtained even in the absence of such a mass term.

This paper is organized as follows.


(Step 1) [Reformulating the Yang-Mills theory in terms of new variables]
In a path-integral quantization for the Yang-Mills theory, we decompose the Yang-Mills field $\mathscr{A}_\mu(x)$ into two pieces $\mathscr{V}_\mu(x)$ and $\mathscr{X}_\mu(x)$, i.e., $\mathscr{A}_\mu(x)=\mathscr{V}_\mu(x)+\mathscr{X}_\mu(x)$, 
and rewrite the action $S_{\rm YM}[\mathscr{A}]$ and the integration measure $[d\mathscr{A}]$ in terms of new variables related to  $\mathscr{V}_\mu(x)$ and $\mathscr{X}_\mu(x)$, according to \cite{Cho80,DG79,FN99} and \cite{KMS06,KMS05,Kondo06,KSM08}. 

(Step 2) [Deriving an effective model by eliminating high-energy modes]
We integrate out $\mathscr{X}_\mu$ field  as the high-energy mode ($p^2 \ge M^2$) with a certain mass scale $M$ of the  field $\mathscr{A}_\mu$. Therefore, the resulting model $S_{\rm YM}^{\rm eff}[\mathscr{V}]$ is written in terms of $\mathscr{V}_\mu(x)$, and is identified with a low-energy effective model for describing the low-energy regime $p^2 \le M^2$. 
A physical reasoning behind this step is explained below. 
The full gauge invariance of the original Yang-Mills theory $S_{\rm YM}[\mathscr{A}]$  is retained also for $S_{\rm YM}^{\rm eff}[\mathscr{V}]$.

However, from the physical viewpoint of clarifying what is the mechanism for confinement, we  modify the step 1 and step 2 as follows. 

(Step 1', 2')
We introduce an antisymmetric tensor field ${}^*B_{\mu\nu}$ of rank 2 \cite{Kondo97,Ellwanger98,Kondo00}, which is interpreted as a composite field of the Yang-Mills field. Then we repeat the same procedures as before to obtain the effective model $S_{\rm YM}^{\rm eff}[\mathscr{V},B]$ by integrating out $\mathscr{X}$ field.

 By integrating out the $B$ field in $S_{\rm YM}^{\rm eff}[\mathscr{V},B]$, another effective theory $\tilde S_{\rm YM}^{\rm eff}[\mathscr{V}]$ is obtained in the gauge-independent manner.

(Step 3) [Converting the Wilson loop to the surface-integral]
In the new formulation using new variables, we can exactly rewrite the Wilson loop operator $W_{C}[\mathscr{A}]$ originally defined in terms of $\mathscr{A}_\mu(x)$ by making use of $\mathscr{V}_\mu(x)$ alone without any reference to $\mathscr{X}_\mu(x)$, according to \cite{DP89,Kondo98b,Kondo08}. 
This fact suggests that $\tilde S_{\rm YM}^{\rm eff}[\mathscr{V}]$ is suitable as a low-energy effective model for quark confinement. 

Up to now, all the results are gauge independent. 
In what follows, we choose a gauge to simplify the calculations.

(Step 4) [Calculating the Wilson loop average to show  area law: quark confinement]
The Wilson loop average $\langle W_{C}[\mathscr{A}] \rangle_{\rm YM}$, i.e., vacuum expectation value of the Wilson loop operator $W_{C}[\mathscr{A}]$ is evaluated by using the effective model $\tilde S_{\rm YM}^{\rm eff}[\mathscr{V}]$ as $\langle W_{C}[\mathscr{A}] \rangle_{\rm YM}^{\rm eff}$.  We show that the Wilson loop average has the area law for sufficiently large loop $C$, leading to the  non-vanishing string tension $\sigma$ in the linear part $\sigma R$ for the static quark-antiquark potential $V(R)$.

Although the area law of the Wilson loop average is obtained also in the original model $S_{\rm YM}^{\rm eff}[\mathscr{V}]$, the modified model $S_{\rm YM}^{\rm eff}[\mathscr{V},B]$ has the advantages:

(Step 5)[Calculating the Schwinger function to show positivity violation: gluon confinement]
The effective gluon propagator for $\mathscr{V}$ of another effective theory $S_{\rm YM}^{\rm eff}[\mathscr{V}]$ (obtained by integrating out the $B$ field  in $S_{\rm YM}^{\rm eff}[\mathscr{V},B]$) has the Gribov-Stingl  form \cite{Stingl86}. 
The Schwinger function calculated from the effective gluon propagator of the Gribov-Stingl  form \cite{Stingl86} exhibits positivity violation suggesting gluon confinement  \cite{KO79,GZ78}.

Thus the derived effective model exhibits both quark confinement (area law) and gluon confinement (positivity violation).

In this paper, we consider only the $SU(2)$ gauge group \cite{KMS06,KMS05,Kondo06} and the extension to $SU(N)$  based on \cite{KSM08}  will be given in a subsequent paper.

(Step 1)
The explicit transformation from the original variables $\mathscr{A}_\mu$ to the new variables $\mathscr{V}_\mu$, $\mathscr{X}_\mu$ are given by
\begin{align}
  & \mathscr{V}_\mu(x)= c_\mu(x)\bm{n}(x) +    ig^{-1} [ \bm{n}(x) , \partial_\mu \bm{n}(x) ]  ,
\nonumber\\
 & \quad\quad\quad\quad  c_\mu(x) :=  \mathscr{A}_\mu(x)  \cdot \bm{n}(x) ,
\nonumber\\
 & \mathscr{X}_\mu(x) =   i g^{-1} [ D_\mu[\mathscr{A}] \bm{n}(x) , \bm{n}(x) ] .
\end{align}
Here $\bm{n}(x)$  is the Lie-algebra $su(2)$-valued field $\bm{n}(x)=n^A(x) T_A $ ($A=1,2,3$) with a unit length, i.e., $n^A(x)n^A(x)=1$.
The so-called color direction field $\bm{n}$ must be obtained in advance as a functional of the original variable $\mathscr{A}_\mu$by solving the reduction condition \cite{KMS06}, e.g., 
$
 [ \bm{n}(x) , D_\mu[\mathscr{A}]D_\mu[\mathscr{A}]\bm{n}(x) ] =0 
$.

The new variable $\mathscr{V}_\mu(x)$ as an   $su(2)$-valued field $\mathscr{V}_\mu(x)=\mathscr{V}_\mu^A(x)T_A$ ($A=1,2,3$) is constructed so that 

(i) $\mathscr{V}_\mu$ has the same gauge transformation as the original field $\mathscr{A}_\mu$, i.e., $\mathscr{V}_\mu(x) \rightarrow \Omega(x) \mathscr{V}_\mu(x) \Omega(x)^\dagger + ig^{-1} \Omega(x) \partial_\mu \Omega(x)^\dagger$ 
and hence its field strength $\mathscr{F}_{\mu\nu}[\mathscr{V}]:= \partial_\mu \mathscr{V}_\nu - \partial_\nu \mathscr{V}_\mu -ig[\mathscr{V}_\mu, \mathscr{V}_\nu ]$ transforms as $\mathscr{F}_{\mu\nu}[\mathscr{V}](x) \rightarrow \Omega(x) \mathscr{F}_{\mu\nu}[\mathscr{V}] \Omega(x)^\dagger$, 
and 

(ii) $\mathscr{F}_{\mu\nu}[\mathscr{V}]$ is proportional to $\bm{n}$, i.e., $\mathscr{F}_{\mu\nu}[\mathscr{V}](x):=\bm{n} (x)G_{\mu\nu}(x)$.

Consequently, $G_{\mu\nu}=\bm{n} \cdot \mathscr{F}_{\mu\nu}[\mathscr{V}]$ is gauge-invariant, since the field $\bm{n}$ is constructed so that it transforms as
$\bm{n}(x) \rightarrow \Omega(x)\bm{n}(x)\Omega(x)^\dagger$. 
Remarkably, $G_{\mu\nu}$ has the same form as the 't Hooft-Polyakov tensor for magnetic monopole:
\begin{equation}
 G_{\mu\nu} 
  =  \partial_\mu c_\nu - \partial_\nu c_\mu +i g^{-1} \bm{n} \cdot [\partial_\mu \bm{n} ,  \partial_\nu \bm{n} ] 
 .
 \label{G}
\end{equation}

(Step 1')
We can introduce a gauge-invariant antisymmetric tensor field $({}^*B)_{\mu\nu}$ of rank 2 by inserting a unity into the path-integral \cite{Kondo97,Ellwanger98,Kondo00}:
\begin{align}
 1 =&  \int \mathcal{D}B \exp \Big[ - \int d^D x
\frac{\gamma}{4} \{   ({}^*B)_{\mu\nu} 
\nonumber\\&
- (\alpha \bm{n} \cdot \mathscr{F}_{\mu\nu}[\mathscr{V}]  - \beta \bm{n} \cdot i g [ \mathscr{X}_\mu , \mathscr{X}_\nu ]) \}^2  \Big] 
 ,
\end{align}
where $*$ is the Hodge dual operation. 
Here  (too many) parameters $\gamma, \alpha, \beta$ are introduced to see effects of each term. 
When $\beta=\gamma^{-1}=\tilde G$ and $\alpha=0$, indeed, $({}^*B)_{\mu\nu}$ is regarded as a collective field for the composite operator $\bm{n} \cdot i g [ \mathscr{X}_\mu , \mathscr{X}_\nu ]$ with the propagator $\tilde G$ obtainable in a self-consistent way  \cite{EW94} according to the Wilsonian renormalization group (RG) \cite{Wetterich93}.  

Then the Euclidean Yang-Mills Lagrangian is rewritten and modified into   
\begin{align}
 & \mathscr{L}_{\rm YM}[\mathscr{V},\mathscr{X},B]
\nonumber\\&
=      \frac{1+\gamma \alpha^2}{4}  G_{\mu\nu}^2
+ \frac{\gamma}{4}  ({}^*B)_{\mu\nu}^2 - \frac{\gamma \alpha}{2}  ({}^*B)_{\mu\nu}   G_{\mu\nu} 
\nonumber\\&
+  \frac{1}{2} \mathscr{X}^{\mu A}  Q_{\mu\nu}^{AB} \mathscr{X}^{\nu B} 
+   \frac{1+\gamma\beta^2}{4} (i g [ \mathscr{X}_\mu , \mathscr{X}_\nu ])^2
 ,
\end{align}
where we have defined 
\begin{align}
Q_{\mu\nu}^{AB} :=& S^{AB} \delta_{\mu\nu} 
+ (2+\gamma\alpha\beta)g\epsilon^{ABC} n^C  G_{\mu\nu} 
\nonumber\\&
- \gamma\beta g\epsilon^{ABC} n^{C}  (*B)_{\mu\nu} 
 ,
 \nonumber\\
 S^{AB} :=& - (D_\rho[\mathscr{V}]D_\rho[\mathscr{V}])^{AB}
  ,
\end{align}
with the covariant derivative $D_\mu$ in the adjoint representation with $\mathscr{V}_\mu:=\mathscr{V}_\mu^C T_C$, $(T_C)^{AB}=if^{ACB}$:
$
    D_\mu^{AB} := \partial_\mu \delta^{AB} -g f^{ABC} \mathscr{V}_\mu^C 
= [\partial_\mu  \mathbf{1} -ig \mathscr{V}_\mu]^{AB}  
$.

[On the effect and the role of the gluon mass term]  
The gluon ``mass term'' for the $\mathscr{X}$ field,
\begin{equation}
 \frac12 M^2 \mathscr{X}_\mu^2 ,
 \label{Mass}
\end{equation}
is gauge invariant in the new formulation \cite{Kondo06}. 
Therefore, we can include this mass term in  calculating  the low-energy effective action.  But we do not introduce this mass term explicitly. 
On the other hand, the inclusion of the gluon mass term for the $\mathscr{V}$ field,
\begin{equation}
 \frac12 m^2 \mathscr{V}_\mu^2 = \frac12 m^2 c_\mu^2 + \frac12 m^2 (\partial_\mu \bm{n})^2 ,
 \label{mass}
\end{equation}
breaks   gauge invariance  and BRST invariance   after taking specific gauges.  However, we can modify the BRST  such that the modified BRST is a symmetry of the Yang-Mills theory with the mass term at the cost of nilpotency. 

(Step2')
We identify $\mathscr{X}_\mu$ with the ``high-energy'' mode in the range $p^2 \in [M^2, \Lambda^2]$ and proceed to integrate out the ``high-energy'' modes $\mathscr{X}_\mu$.
Here $M$ is the infrared (IR) cutoff and $\Lambda$ is the ultraviolet (UV) cutoff as the initial value for the Wilsonian RG. 
In the derivation of our effective model, we neglect quartic self-interactions among $\mathscr{X}_\mu$, i.e., $(i g [ \mathscr{X}_\mu , \mathscr{X}_\nu ])^2$.
\footnote{
However, we can take into account an effect coming from the quartic interaction, which influences our effective model. 
In fact, it is shown \cite{Kondo06,KKMSS05} that the quartic gluon interaction $(i g [ \mathscr{X}_\mu , \mathscr{X}_\nu ])^2$ among  $\mathscr{X}_\mu$ gluons can induce a contribution to the mass term (\ref{Mass}) $\frac12 M^2 \mathscr{X}_\mu^2$  
through a vacuum condensation  of ``mass dimension-2"  (the BRST-invariant version was proposed in \cite{Kondo01}), 
\begin{equation}
 \left< \mathscr{X}_\nu^B(x) \mathscr{X}_\nu^B(x) \right> \not= 0 ,
\end{equation}
which leads to the mass term (\ref{Mass}) with $M^2 \simeq \left< \mathscr{X}_\nu^B(x) \mathscr{X}_\nu^B(x) \right>$ up to a numerical factor.
This result is easily understood by a Hartree-Fock argument. 
This effect is included in the heat kernel calculation through the infrared regularization.
}
In these approximations, we can integrate out $\mathscr{X}_\mu$ by the Gaussian integration and obtain a \textit{gauge-invariant} low-energy effective action $S_{\rm YM}^{\rm eff}[\mathscr{V},B]$ \textit{without mass terms} (\ref{Mass}),(\ref{mass}):
\begin{align}
 & S_{\rm YM}^{\rm eff}[\mathscr{V},B]
\nonumber\\ 
=& \int  \Big[  \frac{1+\gamma \alpha^2}{4}  G_{\rho\sigma}^2
+ \frac{\gamma}{4} ({}^*B)_{\rho\sigma}^2 
- \frac{\gamma \alpha}{2}  ({}^*B)_{\rho\sigma}   G_{\rho\sigma} \Big]
\nonumber\\&
+  \frac{1}{2} \ln \det  Q_{\rho\sigma}^{AB}  
- \ln \det  S^{AB}
 ,
 \label{Seff}
\end{align}
where $\int =\int d^4x $, the functional logarithmic determinant $\frac{1}{2} \ln \det  Q_{\rho\sigma}^{AB}$ comes from   integrating out the $\mathscr{X}$ field, and the last term comes from  the FP-like determinant term \cite{KMS05}  associated with the reduction condition \cite{KMS06}:
\begin{align}
 & \frac{1}{2} \ln \det  Q_{\rho\sigma}^{AB}  
- \ln \det  S^{AB}
\nonumber\\
&=  \int   \frac{g^2\ln  \frac{\mu^2}{M^2}}{(4\pi )^{2}}  \left[
\frac{1}{6} G_{\rho\sigma}^2 -  \frac12 \{ (2+\gamma \alpha\beta) G_{\rho\sigma}-\gamma\beta ({}^*B)_{\rho\sigma} \}^2  
\right]
\nonumber\\&
+  \int    \frac{g^2}{(4\pi )^{2}}  \frac{1}{M^2} \frac{1}{6}   (\partial_\lambda     \{(2+\gamma \alpha\beta) G_{\rho\sigma}-\gamma\beta ({}^*B)_{\rho\sigma} \}     )^2
\nonumber\\&
+ O(\partial^4/M^4)
 .
 \label{lndet}
\end{align}
The gauge fixing is unnecessary in this calculation. Indeed, the resulting effective action (\ref{Seff}) with (\ref{lndet})  is manifestly gauge invariant.
This is one of main results.  
The correct RG $\beta$-function at the one-loop level 
$\beta(g) :=\mu \frac{dg(\mu)}{d\mu}=-b_1g^3+O(g^5)$, $b_1=\frac{22}{3}/(4\pi)^2$
is reproduced in a gauge invariant way when $\gamma\alpha\beta=0$  which follows from e.g. $\alpha=0$ (mentioned above) or $\gamma=0$ (in the case of no $B_{\mu\nu}$ field).

To obtain (\ref{lndet}), we used the heat kernel  to calculate the {\it regularized} logarithmic determinant. Instead of using the standard regulator function $R_{M}$ of the functional RG approach \cite{Wetterich93}, we  restrict the integration range of $\tau$ to $\tau \in [1/\Lambda^2, 1/M^2]$, which corresponds to the momentum-shell integration $p^2 \in [M^2, \Lambda^2]$
\begin{align}
 \ln \det \mathcal{O} =& - \int d^Dx  \lim_{s \rightarrow 0} \frac{d}{ds} \Big[ \frac{ \mu^{2s}}{\Gamma(s)} \int_{1/\Lambda^2}^{1/M^2} d\tau \tau^{s-1} 
\nonumber\\&
\times 
{\rm tr} \left< x | e^{-\tau \mathcal{O}} | x\right>  \Big] 
  ,
\end{align}
where ${\rm tr}$ denotes the trace over Lorentz indices and group indices and $\mu$ is the renormalization scale.
The limit $\Lambda \rightarrow \infty$ should be  understood in what follows. 
These results should be compared with previous works \cite{Kondo97,Ellwanger98,Gies01,Freire02}.
We can show that the mass term  (\ref{Mass})   plays the same role as the IR regulator mentioned above, see \cite{Kondo00}.

(Step 3)
We use a non-Abelian Stokes theorem \cite{DP89,Kondo98b,Kondo08}  to rewrite a non-Abelian Wilson loop operator 
\begin{align}
  W_C[\mathscr{A}]  
:=& {\rm tr} \left[ \mathscr{P} \exp \left\{ ig  \oint_{C} dx^\mu \mathscr{A}_\mu(x) \right\} \right]
 ,
\end{align}
into the area-integral over the surface $\Sigma$ ($\partial \Sigma=C$):
\begin{equation}
 W_C[\mathscr{A}] = \int d\mu_{\Sigma}(\xi) \exp \left[  ig \frac12 \int_{\Sigma: \partial \Sigma=C} G \right] ,
 \label{W[V]}
\end{equation}
where the product measure $d\mu_{\Sigma}(\xi):=\prod_{x \in \Sigma} d\mu(\xi_{x})$ is defined with an invariant measure $d\mu$ on $SU(2)$  normalized as $\int d\mu(\xi_{x})=1$, $\xi_{x} \in SU(2)$. 
In the two-form $G:=\frac12 G_{\mu\nu}(x) dx^\mu \wedge dx^\nu$, $G_{\mu\nu}$ agrees with the field strength  (\ref{G}) under the identification of the color field $\bm{n}(x)$ with a normalized traceless field     
$
 \bm{n}(x) :=  \xi_{x} (\sigma_3/2) \xi_{x}^\dagger 
$.
See also \cite{KSSK11}.

(Step 4)
We proceed to evaluate the Wilson loop average $W(C)=\langle W_{C}[\mathscr{A}] \rangle_{\rm YM}$ by using the effective action $S_{\rm YM}^{\rm eff}[\mathscr{V},B]$, i.e., $\langle W_{C}[\mathscr{A}] \rangle_{\rm YM}
\simeq  \langle W_C[\mathscr{A}] \rangle_{\rm YM}^{\rm eff} $ with the aid of (\ref{W[V]}).

We demonstrate that the simplest way to obtain the area law is to use the low-energy effective action $S^{\rm eff}_{\rm YM}[c,B]$ retained up to terms quadratic and bilinear in  $c$ and $B$:
\begin{align}
 &  S_{\rm YM}^{\rm eff}[c,B]
\nonumber\\ 
=& \int  \Big[  \frac{1+\gamma \alpha^2}{4}  G_{\rho\sigma}^2
+ \frac{\gamma}{4} ({}^*B)_{\rho\sigma}^2 
- \frac{\gamma \alpha}{2}  ({}^*B)_{\rho\sigma}   G_{\rho\sigma} \Big]
\nonumber\\
&+  \int   \frac{g^2\ln  \frac{\mu^2}{M^2}}{(4\pi )^{2}}  \left[
\frac{1}{6} G_{\rho\sigma}^2 -  \frac12 [(2+\gamma \alpha\beta) G_{\rho\sigma}-\gamma\beta ({}^*B)_{\rho\sigma}]^2  
\right]
\nonumber\\&
+  \int    \frac{g^2}{(4\pi )^{2}}  \frac{1}{M^2} \frac{1}{6}   (\partial_\lambda   [(2+\gamma \alpha\beta) G_{\rho\sigma}-\gamma\beta ({}^*B)_{\rho\sigma}]   )^2
\nonumber\\&
+ \int \frac12 m^2 c_\mu^2 + O(\partial^4/M^4)
 ,
 \label{lndet2}
\end{align}
if we include an optional mass term (\ref{mass}).   
\footnote{
The optional mass term (\ref{mass}) is not contained in the original action of Yang-Mills theory.  
But, there is a possibility that such a mass term could be induced in a non-perturbative manner in the effective theory for  the deep infrared region.  
In the full non-perturbative treatment, it is important to avoid the Gribov copies to achieve complete gauge fixing. 
We observe that, in the modified version of the   GZ model \cite{GZ78} called the refined GZ   \cite{RGZ10}, the similar effect to our optional mass term is generated by the horizon term which plays the role of restricting the range of the functional integral to the first Gribov region.  
The horizon term breaks the nilpotent BRST symmetry, so does the optional mass term (\ref{mass}).
This issue is skipped here. 
}

To obtain the propagator or correlation functions, we need to fix the gauge.
For instance, in the  Landau gauge, $\partial^\mu \mathscr{A}_\mu=0$, correlation functions for new variables have been computed on a lattice by numerical simulations using the Monte-Carlo method in \cite{SKKMSI07} based on \cite{KKMSSI05,IKKMSS06}. 
This justifies the identification of $\mathscr{X}_\mu$ as the high-energy mode  negligible in the low-energy regime below $M \simeq 1.2{\rm GeV}$.  

In what follows, we take the unitary-like gauge 
\begin{equation}
 n^A(x) = \delta_{A3} ,
 \label{unitary}
\end{equation}
which reproduces the same effect as taking the MA gauge \cite{tHooft81} in the original Yang-Mills theory.  In this gauge, $\mathscr{X}_\mu^A(x)$ reduces to the off-diagonal component $A_\mu^a(x)$ ($a=1,2$), while $\mathscr{V}_\mu^A(x)$ reduces to the diagonal one $A_\mu^3(x)=a_\mu(x)$, i.e.,
$\mathscr{X}_\mu^A(x)=\mathscr{A}_\mu^a(x) \delta_{Aa} $, $\mathscr{V}_\mu^A(x)=\mathscr{A}_\mu^3(x) \delta_{A3} =c_\mu(x) \delta_{A3} $. In this gauge, the field strength reads
\begin{equation}
 G_{\mu\nu}(x) \rightarrow  
F_{\mu\nu}(x)  :=  \partial_\mu c_\nu(x) - \partial_\nu c_\mu(x), 
\quad
  F=dc  
 ,
\end{equation}
where $d$ denotes the exterior differential. 
The gauge (\ref{unitary}) forces  the color field at each spacetime point to take the  same direction by gauge rotations. Hence the field $c$ contains singularities (of hedge-hog type) similar to the Dirac magnetic monopole after taking the gauge (\ref{unitary}).
Therefore,  $dF=ddc \ne 0$, even if $F=dc$. 
If we do not fix the gauge, such a contribution is contained also in  the part $i g^{-1} \bm{n} \cdot [\partial_\mu \bm{n} ,  \partial_\nu \bm{n} ]$ to make a gauge-invariant  combination $G_{\mu\nu}$ given by (\ref{G}), see \cite{KKMSSI05,IKKMSS06}.

By integrating out the $B$ field, we obtain the effective action $\tilde S^{\rm eff}_{\rm YM}[c]$.
Then we find that the effective propagator $\mathscr{D}_{\rm cc}$ has the Gribov-Stingl form: 
\begin{equation}
 \tilde{\mathscr{D}}_{\rm FF}(p) = p^2 \tilde{\mathscr{D}}_{\rm cc}(p) , \
 \tilde{\mathscr{D}}_{\rm cc}(p) = \frac{1+ d_1 p^2}{ c_0+c_1 p^2+c_2 p^4} 
 ,
  \label{LEET-c-propa2}
\end{equation}
where
$c_0=m^2$,
$c_1=1+ \frac{\gamma\beta^2}{3}  \frac{g^2 }{(4\pi )^{2}}  \frac{m^2}{M^2}$,
$c_2=\frac{g^2}{(4\pi )^{2}}  \frac{1}{M^2}  [(2+\gamma\alpha\beta)^2+(1+\gamma\alpha^2)\gamma\beta^2+2(2+\gamma\alpha\beta)\gamma\alpha\beta]/3$,
and
$d_1= \frac{\gamma\beta^2}{3}  \frac{g^2}{(4\pi )^{2}}  \frac{1}{M^2}$.
The precise values of the parameters $m, \gamma, \alpha, \beta$ and $M$ are to be determined by the functional RG \cite{Wetterich93} following \cite{Kondo10}, which is a subject of   future study.

In the unitary-like gauge (\ref{unitary}) the Wilson loop operator is reduced to
\begin{equation}
    W_C[F] 
=   \exp \left[  ig \frac12 \int_{\Sigma: \partial \Sigma=C} F \right]
= \exp \left[  ig  \frac12(\Theta_{\Sigma},F)  \right] ,
\end{equation}
where $\Theta_{\Sigma}$ is  the vorticity tensor defined by 
$
 \Theta^{\mu\nu}_\Sigma(x) = \int_{\Sigma} d^2S^{\mu\nu}(x(\sigma)) \delta^D(x-x(\sigma)) ,
$
which  has the support on the surface $\Sigma$ whose boundary is the loop $C$.
Here $( \cdot , \cdot )$ is the $L^2$ inner product for two differential forms:
$
(\Theta_{\Sigma},F)  =   \int d^Dx \frac12 \Theta_\Sigma^{\mu\nu}(x) F_{\mu\nu}(x)
$.
  By integrating out $B$, we obtain the effective model 
$S^{\rm eff}_{\rm YM}[c] = \frac{1}{2} \left(  c, \mathscr{D}_{\rm cc}^{-1} c \right)= \frac{1}{2} \left(  F, \mathscr{D}_{\rm FF}^{-1} F \right)$ 
  in terms of $c$ or $F$. Then the Wilson loop average $W(C)$ is evaluated by integrating out $F=dc$:
\begin{align}
 W(C) = \exp \left[
-\frac18 g^2 ( \Theta_{\Sigma}, \mathscr{D}_{\rm FF}  \Theta_{\Sigma}) 
 \right] ,
\end{align}
where $\mathscr{D}_{\rm FF}=\varDelta\mathscr{D}_{\rm cc}$ and its Fourier transform $\tilde{\mathscr{D}}_{\rm FF}(p)=p^2\tilde{\mathscr{D}}_{\rm cc}(p)$.
For concreteness, we choose $\Theta_{\Sigma}$ for a planar surface bounded by a rectangular loop $C$ with side lengths $T$ and $R$ in the $x_3-x_4$ plane.  Then we find that the Wilson loop average has the area law for  large $R$ 
\begin{equation}
W(C) \sim \exp [-\sigma RT] ,
\end{equation} 
with the string tension given by the formula:
\begin{equation}
  \sigma = \frac18 g^2 \int_{p_1^2+p_2^2 \le M^2} {dp_1dp_2 \over (2\pi)^2}   
\tilde{\mathscr{D}}_{\rm FF}(p_1,p_2,0,0) >0
  ,
  \label{st}
\end{equation}
where the momentum integration is restricted to the two-dimensional space (the dimensional reduction by two \cite{Kondo98a}) and is cutoff at the upper limit $M$. 
A positive and finite string tension $0< \sigma < \infty$ 
 follows from the condition of no real poles in the effective gluon propagator $\tilde{\mathscr{D}}_{\rm cc}(p)$ in the Euclidean region, $0< \tilde{\mathscr{D}}_{\rm FF}(p)=p^2 \tilde{\mathscr{D}}_{\rm cc}(p) < \infty$, which is connected to gluon confinement below.
This is another of main results.

According to numerical simulations in MA gauge \cite{AS99,BCGMP03,MCM06}, the diagonal gluon propagator is well fitted to the form (\ref{LEET-c-propa2}): e.g. \cite{MCM06} give
$c_0= 0.064(2){\rm GeV}^2$,
$c_1= 0.125(9)$,
$c_2= 0.197(9){\rm GeV}^{-2}$,
$d_1= 0.13(1){\rm GeV}^{-2}$,
and 
$M \simeq 0.97{\rm GeV}$,
where $M$ is the mass of  off-diagonal gluons  obtained  in the MA gauge. This value of $M$ is a little bit smaller than the values of other groups \cite{AS99,BCGMP03}.
This indeed leads to a good estimate for the string tension  
$\sigma \simeq (0.5{\rm GeV})^2$ according to  (\ref{st}) for    $\alpha(\mu)=g^2(\mu)/(4\pi) \simeq 1.0$ at $\mu=M$. 
The next task is to study how the results are sensitive to  the deep infrared behavior of the diagonal gluon propagator (\ref{LEET-c-propa2})  and the actual value of $M$ for the off-diagonal gluon propagator.

The Gribov-Stingl form is obtained only when $c_0 \ne 0$ (i.e., $m \ne 0$) and $d_1 \ne 0$ ($B_{\mu\nu}$ is included).
Even in the limit $m^2  \rightarrow 0$ $(c_0 \rightarrow 0)$, the area law survives according to (\ref{st}), provided that  
$\tilde{\mathscr{D}}_{\rm FF}(p)$ remains positive and finite: 
$ 
 \tilde{\mathscr{D}}_{\rm FF}(p) 
\rightarrow 
    \frac{ 1+ d_1 p^2}{  c_1 +c_2 p^2} 
$, 
while  $\tilde{\mathscr{D}}_{\rm cc}(p)$ behaves unexpectedly as 
$
 \tilde{\mathscr{D}}_{\rm cc}(p) \rightarrow    \frac{1+ d_1 p^2}{p^2(c_1+c_2p^2)}
$.
Hence, we  argue that it does not matter to quark confinement whether $m=0$ or $m \ne 0$.

(Step5)
The positivity violation is examined. 
In the case of $c_2 = 0$, there is no positivity violation, as far as $c_0/c_1>0$. 
In the case of $c_2 \ne 0$, $\tilde{\mathscr{D}}_{\rm cc}(p)$ has a pair of complex conjugate poles at $p^2=z$ and $p^2=z^*$,  
$
  z := x +i y
$,
$
 x := - c_1/(2c_2)
$,
$
 y := \sqrt{c_0/c_2 - \left( c_1/(2c_2) \right)^2}
$. 
We find that the Schwinger function  
$
\Delta(t) 
  := \int_{-\infty}^{+\infty} \frac{dp_4}{2\pi}  e^{ip_4 t} \tilde{\mathscr{D}}_{\rm cc} (\bm{p}=0,p_4)
$
is oscillatory in $t$ and 
is negative over finite intervals in the Euclidean time $t >0$:
\begin{align}
  \Delta(t) 
&=   \frac{1}{2c_2 |z|^{3/2}\sin (2\varphi)} e^{-t |z|^{1/2} \sin \varphi } [ \cos (t|z|^{1/2} \cos \varphi -\varphi ) 
\nonumber\\&
   + d_1 |z| \cos ( t|z|^{1/2} \cos \varphi +\varphi )  ] 
 ,
\end{align}
where
$ 
z = |z|e^{2i\varphi }
$
with
$
|z|  = \left( c_0/c_2  \right)^{1/2}
$,
$
  \cos (2\varphi) 
= - \sqrt{ c_1^2/(4c_0c_2)} 
$,
and
$
  \sin (2\varphi) 
= \sqrt{1- c_1^2/(4c_0c_2)}
$.
 Therefore, the reflection positivity is violated for the gluon propagator (\ref{LEET-c-propa2}), as long as $0< \frac{c_1^2}{4c_0 c_2} <1$, irrespective of   $d_1$. 
When $c_0 = 0$ (or $m=0$),  
\begin{equation}
  \Delta(t) 
=  -   \frac{t}{2c_1}     -  \frac{1}{2c_1} \sqrt{\frac{c_2}{c_1} } \left( 1-\frac{c_1}{c_2} d_1 \right) e^{-t \sqrt{\frac{c_1}{c_2}}} 
 .
\end{equation}
Hence, the special case $c_0=0$ also violates the positivity, if $c_1 >0$ and $c_2 >0$.
Thus the diagonal gluon in  the MA gauge can be confined.


In summary, we have derived a novel low-energy effective model of the $SU(2)$ Yang-Mills theory without fixing the original gauge symmetry. 
It is remarkable that the effective model respects the  $SU(2)$ gauge invariance of the original Yang-Mills theory, which allows one to take any gauge fixing in computing physical quantities of interest in the low-energy region. 
The effective gluon propagator belongs to the Gribov-Stingl form.
In MA gauge, the model exhibits both quark confinement and gluon confinement simultaneously in the sense that the Wilson loop average satisfies the area law (i.e., the linear quark-antiquark potential) and that the Schwinger function violates  reflection positivity.   
More results and full details will be given in a subsequent paper.



{\it Acknowledgements}\ ---
This work is  supported by Grant-in-Aid for Scientific Research (C) 21540256 from Japan Society for the Promotion of Science
(JSPS).


\end{document}